\documentclass[twocolumn,showpacs,prl,amsmath,amssymb]{revtex4}
\usepackage{graphicx}
\usepackage{bm}
\begin{document}

\title{Radiation-Induced ``Zero-Resistance State''
at Low Magnetic Fields and 
near Half Filling of the Lowest Landau Level}
\author{K. Park}
\affiliation{Condensed Matter Theory Center, 
Department of Physics, University of Maryland,
College Park, MD 20742-4111}
\date{\today}

\begin{abstract}
We theoretically predict the radiation-induced 
``zero-resistance state'' near half filling 
of the lowest Landau level,
which is caused by the photon-assisted transport
in the presence of oscillating density of states due to
composite fermion Landau levels,
and is analogous to the radiation-induced 
``zero-resistance state'' of electrons at low magnetic fields.
Based on a non-perturbative theory,
we show that the radiation field does not  
break the bound state of electrons and flux quanta, 
i.e. composite fermions.
Our prediction is independent of 
the power and frequency of radiation field.
\end{abstract}
\pacs{}
\maketitle

Recently discovered states of a two-dimensional (2D) electron system 
with vanishingly small resistance 
in the presence of microwave radiation 
at low magnetic fields \cite{Mani,Zudov1}
reveal a fundamentally new non-equilibrium state of matter.
This is so not just because of the apparent zero resistance 
(which is the main reason why there has been intense
attention to this problem), 
but also because of magnetoresistance oscillation
which eventually leads to the vanishingly small resistance.
This radiation-induced magnetoresistance oscillation
sheds new light on the interaction 
of radiation and 2D electron system.

It is generally believed that
the zero-resistance state 
is due to
an instability caused by photon-assisted transport 
in the presence of an oscillating density of states,
which leads to negative conductivity
for sufficiently large radiation power.
In spite of this overall consensus,
it is not clear whether previous theoretical frameworks
adequately treated the rather strong interaction between 
the radiation field and the 2D electron systems 
under a magnetic field.
In this article,
we develop a non-perturbative theory which treats
the interaction between radiation and electrons exactly
(though the scattering with impurities is treated perturbatively).

Using this non-perturbative theory,
we would like to make two predictions.
First, it is predicted that 
there is a {\it resonantly enhanced} magnetoresistance oscillation
near $\omega \simeq \omega_c$
which may reveal higher harmonic oscillations.
Second, we predict similar ``zero-resistance states'' 
near half filling of the lowest Landau level ($\nu=1/2$) 
using the composite fermion theory \cite{Jain}.
In particular, 
we show that the radiation field does not  
break the bound state of electrons and flux quanta, 
i.e. composite fermions,
regardless of the power and frequency of the radiation.
Finally, 
note that we will not address the question how the negative 
resistance leads to the zero resistance \cite{Millis}. 
Instead, we will focus
on the nature of the magnetoresistance oscillations
at low magnetic fields and near $\nu=1/2$.

Let us begin by writing the Hamiltonian 
in the presence of microwave radiation. 
The radiation wavelength is long enough so that the electric field
is coherent and uniform in space over the whole sample.
We will choose the Landau gauge with $\textbf{A}=B(0,x)$
in which the momentum in $y$ direction ($k_y$)
is a good quantum number.
Then, for eigenstates with $k_y = -X_j/l_B^2$ 
(where $l_B=\sqrt{\hbar c/eB}$ is the magnetic length), 
the single particle Hamiltonian is written:
\begin{eqnarray}
H &=& \frac{p_x^2}{2M}+ \frac{1}{2}M\omega_c^2 (x-X_j)^2
-F(t)(x-X_j) -F(t)X_j
\nonumber \\
&=& H_F(p_x,x-X_j) -F(t)X_j
\label{hamil}
\end{eqnarray}
where 
$F(t)=eE\cos{\omega t}$,
$E$ is the electric field,
and $H_F(p_x,x-X_j)$ is the Hamiltonian of
a time-dependent forced harmonic oscillator centered at $x=X_j$.

The time-dependent forced harmonic oscillator 
described by $H_F(p_x,x)$
can be solved exactly
for general $F(t)$ without using any perturbation theory: 
the analytic solution is given as follows:
\begin{equation}
\psi(x,t) = \chi(x-\xi(t),t) e^{i \theta(x,t)}
\label{psi_FHO}
\end{equation}
where
$\theta(x,t) = \frac{M}{\hbar}\frac{d \xi(t)}{d t} (x-\xi(t))
+\frac{1}{\hbar}\int^t_0 L dt'$.
$\chi(x,t)$ satisfies the Schr\"{o}dinger equation of
the unforced harmonic oscillator, and can be taken to be 
either the usual number eigenstate or 
any other non-stationary solution such as a coherent state.
The center position $\xi(t)$ satisfies the classical motion of
a forced harmonic oscillator: 
$M \frac{d^2 \xi}{d t^2}+ M \omega_c^2  \xi = F(t)$
where $\omega_c = eB/mc$.
$L$ is the classical Lagrangian:
$L= \frac{M}{2}(\frac{d \xi}{d t})^2 -\frac{M\omega_c^2}{2}\xi^2
+F(t) \xi$.
The key point is that,
other than the phase factor, 
the analytic solution in Eq.~(\ref{psi_FHO})
is precisely the same wavefunction
as an unforced harmonic oscillator
with center position displaced by $\xi(t)$.
Since $\xi(t)$ is 
just the classical solution of the forced harmonic oscillator, 
it is independent of any quantum numbers.

The analytic solution of 
the Hamiltonian in Eq.~(\ref{hamil}) 
is then given by:
\begin{eqnarray}
\phi_{n,j}(x,t) &=& \psi_{n}(x-X_j,t) 
\exp{\left(\frac{i}{\hbar} X_j \int^t_0 dt' F(t') \right)} 
\nonumber \\
&=& \chi_{n}(x-X_j-\xi(t),t) e^{i \theta(x,t)}
\nonumber \\
&\times& 
\exp{\left(\frac{i}{\hbar} X_j 
\left\{
\int^t_0 dt' F(t') 
-M\frac{d \xi(t)}{d t}
\right\}
\right)} 
\label{phi}
\end{eqnarray}
where $n$ is the Landau level index.
The phase factor
$\exp{\left(\frac{i}{\hbar} X_j \int^t_0 dt' F(t') \right)}$
is due to the last term in Eq.~(\ref{hamil}).
Strictly speaking, 
the energy is not conserved 
in the presence of radiation.
However,
the Landau level index $n$ is 
an invariant quantum number
which is connected to 
the usual Landau level index 
of unforced harmonic oscillator.
It is easy to understand 
the meaning of invariance if one considers
the situation where
an electron is in the $n$-th Landau level initially, 
and the radiation field is turned on at $t=0$.
The electron will stay in the exact state in Eq.~(\ref{phi})
with the same Landau level index $n$ at $t>0$.

\begin{figure}
\includegraphics[width=3.2in]{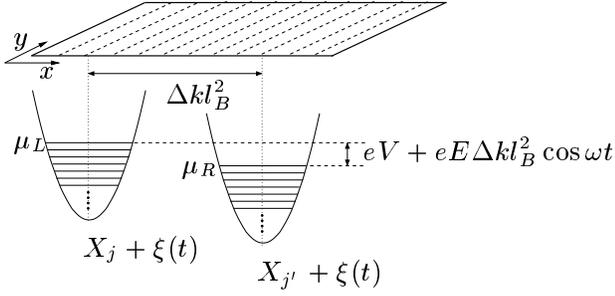}
\caption{Schematic diagram of photon-assisted transport.
Center coordinates ($X_j$) of momentum eigenstate 
with momentum $k_y$ in the Landau gauge
are denoted as dashed lines in 2D plane in top panel;
$X_j=-k_y l_B^2$.
Each momentum eigenstate is translated 
by exactly the same displacement $\xi(t)$
in the presence of radiation field.
Schematically shown here is the scattering between the eigenstate
with $X_j$ and $X_{j'}=X_j+\Delta k l^2_B$ where
$\Delta k$ is the mean momentum transfer 
due to impurity scattering.
Though the transmission coefficient is
determined predominantly by the scattering channel
in which the overlap between 
wavefunctions in left and right sides is the highest,
two parabolic confinement potentials are
drawn to be separated far for the sake of clarity.
\label{fig1}}
\end{figure}

Because the orbital part of exact solution, 
$\phi_{n,j}(x,t)$, in Eq.~(\ref{phi})
is quantized in the exactly same way 
as eigenstates in the absence of radiation,
the density of states should be identical to
that without a radiation field,
if there were no time-dependent phase factor.
In order to understand how the density of states should be
modified because of the time-dependent phase factor,
let us rewrite $\phi_{n,j}(x,t)$ as follows:
\begin{eqnarray}
\phi_{n,j}(x,t) &=& \chi_{n}(x-X_j-\xi(t),t) e^{i \theta(x,t)}
\nonumber \\
&\times& 
\exp{\left(i \frac{eE}{\hbar} 
X_j 
\left(
\frac{1}{\omega} +\frac{\omega}{\omega^2_c-\omega^2}
\right)
\sin{\omega t}
\right)} 
\nonumber \\
 &=& \chi_{n}(x-X_j-\xi(t),t) e^{i \theta(x,t)}
\nonumber \\
&\times&
\sum^{\infty}_{m=-\infty} J_m
\left(
\frac{eE}{\hbar} X_j 
\left(
\frac{1}{\omega} +\frac{\omega}{\omega^2_c-\omega^2}
\right)
\right) e^{im\omega t}
\nonumber \\
\label{expansion}
\end{eqnarray}
where  
the Jacobi-Anger expansion is used:
$e^{iz\sin{\phi}}= \sum_m J_m(z) e^{im\phi}$.
Note that
$\theta(x,t)$ is independent of the center coordinate $X_j$.
Following Tien and Gordon \cite{Tien-Gordon},
Eq.~(\ref{expansion}) can be viewed in such a way that
the wavefunction contains components
which have energies, 
$\epsilon_n, \epsilon_n\pm\hbar\omega, 
\epsilon_n\pm 2\hbar\omega, \cdots, \epsilon_n\pm m\hbar\omega$, etc. ,
and each component is weighted by $J_m$. 
Therefore,
the density of states
should also contain contributions from each energy component: 
\begin{eqnarray}
\rho ' (\epsilon) = \sum_m 
\rho (\epsilon +m\hbar\omega)
J^2_m
\left(
\frac{eE}{\hbar} X_j 
\left(
\frac{1}{\omega} +\frac{\omega}{\omega^2_c-\omega^2}
\right)
\right)
\label{effective_dos}
\end{eqnarray}
where $\rho(\epsilon)$
is the density of states without radiation.
Note that, 
though $\theta(x,t)$ is dependent on time (and position),
it does not affect any of the energy components
because it does not depend on 
the center coordinates $X_j$.
So one can get rid of $\theta(x,t)$
via a proper gauge transformation
without affecting physical consequences.

Now, let us discuss 
the effect of a modified density of states 
in the context of transport which is formulated
via the center-migration theory \cite{Ando,Zudov2}.
A schematic diagram is shown in Fig.\ref{fig1}.
In the Landau gauge with $\textbf{A}=B(0,x)$, 
electrons travel freely along the $y$ direction,
while confined in the $x$ direction,
which causes the Hall effect.
Longitudinal resistance (in the $x$ direction) appears 
when electrons are scattered by impurities 
so that there is a momentum transfer $\Delta k$. 
This momentum transfer is equivalent to 
a spatial jump in the center coordinates along the $x$ direction
by the distance of $|\Delta X_j| = |\Delta k| l^2_B$. 
The probability of jumps in the center coordinate is 
proportional to the overlap between
the wavefunctions of initial and final 
center coordinates:
\begin{equation}
P(\Delta X_j)=  \left| 
\int^{\infty}_{-\infty} dx \;\chi_n(x-X_j) \chi_{n'}(x-X_{j'})
\right|^2 .
\label{overlap}
\end{equation}
The Landau level index of the highest occupied level 
is sufficiently large for low magnetic fields:
$n \simeq n' \simeq 50$ for systems studied in experiments.
So,
$\chi_n(x)$ is sharply peaked near the edges of 
the wave packet around $x = \pm \sqrt{2n+1} l_B$.
Therefore,
$P(\Delta X_j)$ is also sharply peaked around
$\Delta X_j \simeq 2\sqrt{2n+1} l_B$.
Assuming that the transport occurs predominantly 
through this channel of momentum transfer
$|\Delta k| = |\Delta X_j| /l^2_B \simeq 2\sqrt{2n+1}/ l_B$,
we can write the following formula for current
in the presence of an external bias $eV$:
\begin{eqnarray}
I &=& e D  
\sum^{\infty}_{m=-\infty} J^2_m (\alpha)  
\nonumber \\
&\times& \int d\epsilon \rho(\epsilon) \rho(\epsilon+m\hbar\omega+eV)
[ f(\epsilon) -f(\epsilon +m\hbar\omega +eV) ]
\nonumber \\
\end{eqnarray}
where 
$\alpha =
\frac{eE}{\hbar} \Delta X_j 
\left(
\frac{1}{\omega} +\frac{\omega}{\omega^2_c-\omega^2}
\right)$,
$f(\epsilon)$ is the usual Fermi-Dirac distribution function and
$D$ is the transmission coefficient
(which is proportional the overlap in Eq.~(\ref{overlap}))
between two eigenstates separated by the distance  
$\Delta X_j \simeq 2\sqrt{2n+1} l_B$.
Then, the zero-bias conductivity 
$\sigma (\equiv \frac{d I}{d V}|_{V=0})$
can be written as follows:
\begin{eqnarray}
\sigma = e^2&D&   
\sum^{\infty}_{m=-\infty} 
J^2_m (\alpha)
\nonumber \\
&\times& 
\int d\epsilon 
\big\{
-f'(\epsilon+m\hbar\omega) 
\rho(\epsilon) \rho(\epsilon+m\hbar\omega)
\nonumber \\
&+&[ f(\epsilon) -f(\epsilon +m\hbar\omega) ]
\rho(\epsilon) \rho'(\epsilon+m\hbar\omega)
\big\}
\label{sigma}
\end{eqnarray}
Eq.~(\ref{sigma}) is similar to the formula obtained by 
Shi and Xie\cite{Xie}. 
However, it is important to note that
there is a significant difference 
in the argument of Bessel function
The argument shows a resonance behavior 
which was not obtained
in previous perturbative theories. 
Consequences of this resonant behavior
will be discussed in detail later.

Until now, the effect of impurities
has been included only in a way that
they cause the scattering between
states with different center coordinates.
Another important effect of impurities is
the broadening of Landau levels so that
the density of states is not 
a series of delta functions.
Following Ando {\it et al.} \cite{Ando},
we assume the following form 
for the density of states (in the absence of radiation):
\begin{eqnarray}
\rho(\epsilon) = \rho_0 
\left(
1 - \lambda \cos{\frac{2\pi\epsilon}{\hbar\omega_c}}
\right)
\label{dos}
\end{eqnarray}
where 
$\lambda = 2 e^{-\pi/\omega_c \tau_f}$
and $\tau_f$ is the relaxation time.

It is now straightforward to 
compute the conductivity by plugging 
the density of states in Eq.~(\ref{dos}) into Eq.~(\ref{sigma}).
As it can be shown by full integration,
the photon-assisted conductivity 
basically gives rise to a slow oscillation
as a function of inverse of magnetic field:
$\propto \sin({2\pi m \omega/\omega_c})$
with $m$ an integer,
while the usual Shubnikov-de Haas oscillation
is a fast oscillation 
$\propto \sin({2\pi \mu/\hbar\omega_c})$
with $\mu$ the chemical potential.
One obtains a separation between 
the fast Shubnikov-de Haas oscillations and 
the slow oscillations in radiation-induced conductivity
when 
(i) the chemical potential energy is much larger 
that the temperature: $\mu/k_B T \gg 1$, and
(ii) the temperature is not too low, 
compared to the Landau level spacing:
$k_B T/\hbar\omega_c \sim {\cal O}(1)$. 
Note that
the regime of our interest,
$\hbar\omega_c \lesssim k_B T \ll \mu$,
is consistent with 
the parameter range of experiments.

In this regime,
the contribution of photon-assisted conductivity
can be written as follows:
\begin{eqnarray}
\frac{\sigma -\sigma_{\textrm{SdH}}}{\sigma_0}
&=&\frac{\lambda^2}{2} 
\sum_{m=-\infty}^{\infty}
J^2_m (\alpha)
\nonumber \\
&\times&
\left[
\cos{\left(2\pi m \frac{\omega}{\omega_c}\right)}
-2\pi m \frac{\omega}{\omega_c} 
\sin{\left(2\pi m \frac{\omega}{\omega_c}\right)}
-1
\right]
\nonumber \\
\label{final_sigma}
\end{eqnarray}
where
$\sigma_0 = e^2 D \rho^2_0$
and $\sigma_{\textrm{SdH}}$ is the conductivity 
in the absence of radiation, 
which is nothing but the conductivity due to
the usual Shubnikov-de Haas effect.
In other words,
$\sigma_{\textrm{SdH}}$ can be defined as
the conductivity when $E=0$.
Also, as discussed previously, 
$\alpha =
\frac{eE}{\hbar} \Delta X_j 
\left(
\frac{1}{\omega} +\frac{\omega}{\omega^2_c-\omega^2}
\right)$ with
$\Delta X_j \simeq 2\sqrt{2n+1} l_B$.
There are several points to be emphasized.

First,
the $m$-th harmonic 
of photon-assisted conductivity oscillation
is weighted by $J^2_m(\alpha)$.
When $\alpha \ll 1$,
the only important term is the zeroth harmonic 
which is just a constant and can be absorbed into
the conductivity in the absence of radiation.
As $\alpha$ increases, 
weights of other harmonics become important
because the sum of all weights is unity, 
i.e. $\sum_{m=-\infty}^{\infty} J^2_m(\alpha) =1$,
and the weight of zeroth harmonic decreases 
(though oscillatory).
In particular, 
$\alpha$ is resonantly enhanced
when $\omega$ is very close to $\omega_c$ \cite{comment_resonance}.
Therefore,
it is predicted that,
in addition to the first harmonic ($m=1$) oscillation,
there should be prominent oscillations 
of other harmonics ($m\geq2$)
near $\omega \simeq \omega_c$.
This additional harmonic effect 
may have already been observed 
in experiments \cite{Zudov1,Zudov3}.


Second,
Eq.~(\ref{final_sigma}) explicitly shows that
the conductivity in the presence of radiation is identical to
that without radiation if $\omega/\omega_c$ is an integer.  
In fact, this property is generally true
for any periodic density of states.
Therefore,
it is a much more robust feature
that the position of ``zero-resistance state''\cite{Xie,Sachdev} .

Third, 
superficially,
temperature dependence is absent in Eq.~(\ref{final_sigma}).
The temperature dependence of radiation-induced resistance oscillations
originates from the $T$ dependence of scattering,
either 
through the broadening of the density of states 
(via the relaxation time $\tau_f$ in Eq.~(\ref{dos}))
or
through the $T$ dependence of
scattering between states with different center coordinates. 
Similar arguments were given by 
Shi and Xie \cite{Xie}, and
Durst {\it et al.} \cite{Sachdev}.

Fourth, 
Eq.~(\ref{final_sigma}) describes 
the longitudinal conductivity $\sigma_{xx}$
which is oscillatory.
On the other hand, our theoretical framework naturally predicts that
the Hall conductivity $\sigma_{xy}$ in the presence of radiation
is identical to that without radiation.
Therefore, 
$\rho_{xy} = \sigma_{xy}/(\sigma^2_{xx}+\sigma^2_{xy})$
should  be roughly equal to the $\rho_{xy}$
without radiation if $\sigma_{xx} \ll \sigma_{xy}$,
but may have some signature of oscillatory behavior
for intermediate values of $\sigma_{xx}$ \cite{R_xy}.
Also, 
because
$\rho_{xx}=\sigma_{xx}/(\sigma^2_{xx}+\sigma^2_{xy})
\simeq \sigma_{xx}/\sigma^2_{xy}$ for $\sigma_{xx} \ll \sigma_{xy}$,
the longitudinal resistivity
is proportional to the longitudinal conductivity.

Now we make connection between the states at low magnetic fields and
near $\nu=1/2$.
In particular,
we would like to predict the radiation-induced
magnetoresistance oscillation  
near half filling of the lowest Landau level,
based on the composite fermion (CF) theory \cite{Jain}.
The random phase approximation (RPA) of 
fermionic Chern-Simons gauge theory \cite{CS1,CS2,HLR}
predicts the following, simple relationship
between the resistance of systems of strongly interacting electrons
at filling factor $\nu=\nu^*/(2p\nu^*+1)$ ($p$ is an integer)
and that of weakly interacting composite fermions
at  $\nu^*$:
\begin{eqnarray}
\rho = \rho_{\textrm{CF}}+\rho_{\textrm{CS}}
\end{eqnarray}
where 
$\rho_{CF}$ is the resistivity matrix of composite fermions and
\begin{eqnarray}
\rho_{\textrm{CS}} = 
\frac{h}{e^2}
\left[ 
\begin{array}{cc}
0 & 2p \\
-2p & 0 
\end{array}
\right],
\end{eqnarray}
where $2p$ is the number of flux quanta captured by 
composite fermions.
In particular,
the composite fermion theory tells us that, for $p=1$,
there is a mapping between the states in the vicinity of $\nu=1/2$
and the states at low magnetic fields, i.e. $\nu^* \gg 1$.
So, if the longitudinal resistance of weakly interacting
fermion systems at low magnetic fields 
vanishes in the presence of microwave radiation,
the composite fermion theory
predicts a vanishing longitudinal resistance for
systems of strongly interacting electrons near $\nu=1/2$,
provided that the radiation field does not break
composite fermions
which is a bound state of electron and flux quanta. 
Now we show that the radiation field
does not break the composite fermions
(as long as the electric field is coherent and uniform in space).

We begin by showing that
the ground state in the presence of long-wavelength radiation
is identical to
the ground state without radiation (up to the center-of-mass motion)
regardless of the inter-electron interaction.
To this end,
let us write the general Hamiltonian in a perpendicular magnetic field
with inter-electron interaction as well as radiation:
\begin{eqnarray}
H = \frac{1}{2M} \left(\textbf{p} - \frac{e}{c}\textbf{A}\right)^2
+\sum_{i<j} V(r_{ij})
+H_{\textrm{rad}} ,
\end{eqnarray}
where 
the radiation Hamiltonian
$H_{\textrm{rad}}=-F(t) \sum_i x_i =-F(t) x_{\textrm{cm}}$
with $x_{\textrm{cm}}$ being the center-of-mass coordinate,
$V(r_{ij})$ is the interelectron interaction,
and $r_{ij}$ is the distance between the $i$-th and $j$-th electron.
Since the kinetic energy term in the Hamiltonian is 
quadratic in $\textbf{r}$ and $\textbf{p}$ and is linear in 
$\textbf{L} = \textbf{r}\times\textbf{p}$,
the center-of-mass and relative coordinates decouple,
which is known as the generalized Kohn theorem \cite{Kohn}.
Consequently, 
the interaction with radiation in the long-wavelength limit 
can not change the state of relative motion
since $H_{\textrm{rad}}$ depends only on
the center-of-mass coordinate.


A more explicit proof in the lowest Landau level
can be obtained as follows. 
As shown previously, 
the Landau level index 
is invariant 
in the presence of radiation so that electrons stay 
in the exact states in Eq.~(\ref{phi})
with the same Landau level index $n$
(and also with the same center coordinate index $X_j$,
though the center position becomes $X_j+\xi(t)$).
So,
there is a one-to-one correspondence between
the basis states of the lowest Landau level
with and without radiation.
More importantly,
interaction with radiation 
does not change the shape or the relative distance 
between the basis states. 
Therefore, 
the Coulomb matrix elements do not change, 
nor does the relative-coordinate part of the ground state
which originates from Coulomb correlation.

Now, it is important to note that
Coulomb correlation in the lowest Landau level
manifests itself as flux attachment because
flux quanta are actually zeroes of the many-body wavefunction,
and electrons try to minimize their Coulomb energy cost
by forming a bound state with zeroes.
Also, note that, after the flux attachment, 
electron systems near $\nu=1/2$ are mapped onto
weakly interacting systems of composite fermions
which form effective Landau levels 
with low effective magnetic field
$B^*=B-B_{1/2}$ where $B_{1/2}$ is the magnetic field at $\nu=1/2$.
Therefore, our theory of photon-assisted transport at low fields
can be directly applied to the CF states near 
$\nu=1/2$ \cite{comment_rhoxx}.

We may also establish the period of radiation-induced
magnetoresistance oscillation near $\nu=1/2$.
Remember that, at low fields, the megnetoresistance oscillation 
is periodic as a function of $\omega/\omega_c$.
In the case of composite fermion systems,
the cyclotron frequency should be modified to
the effective cyclotron frequency of composite fermions:
\begin{eqnarray}
\omega^*_c = \frac{e B^*}{m^* c} = \frac{e}{m^* c}(B-B_{1/2})
\end{eqnarray}
where the CF mass $m^*$ is given as follows:
\begin{eqnarray}
m^* = C \frac{\varepsilon\hbar^2}{e^2} \sqrt{\frac{eB}{\hbar c}}
= C \frac{\varepsilon\hbar^2}{e^2}\frac{1}{l_B}
\end{eqnarray}
where $\varepsilon$ is the dielectric constant and
$C \simeq 3.0$ \cite{HLR,Du,JainCFmass} 
is a dimensionless constant.
For a typical value of magnetic field $B \simeq 10$ T for $\nu=1/2$,
$m^*$ is roughly 3 to 4 times larger than the band mass of
electron in GaAs \cite{comment_CFmass}.
Therefore, overall,
the radiation-induced megetoresistance oscillation
near $\nu=1/2$ will have 3 to 4 times shorter period
as a function of $1/(B-B_{1/2})$
for a given value of radiation frequency $\omega$.
However, it is important to 
note that the CF mass itself depends weakly on the magnetic field
so that there will be a small correction to the period of oscillation:
$\omega/\omega_c^* \propto \frac{\sqrt{B_{1/2}+B^*}}{B^*}$.
Finally, note that the measurement of radiation-induced magnetoresistance 
can be regarded as a way of measuring the effective mass of 
composite fermion.

We acknowledge helpful conversations with 
A. Kaminski, V. W. Scarola, Juren Shi, and X. C. Xie.  
This work was supported by ARDA.




\begin{thebibliography}{}

\bibitem{Mani} R. G. Mani {\it et al.}, Nature (London) {\bf 420},
6464 (2002).


\bibitem{Zudov1} M. A. Zudov, R. R. Du, L. N. Pfeiffer, and K. W. West,
Phys. Rev. Lett. {\bf 90}, 046807 (2003).



\bibitem{Jain} J. K. Jain,
Phys. Rev. Lett. {\bf 63}, 199 (1989).


\bibitem{Millis} A. V. Andreev, I. L. Aleiner, and A. J. Millis,
Phys. Rev. Lett. {\bf 91}, 056803 (2003).


\bibitem{Tien-Gordon} P. K. Tien and J. P. Gordon,
Phys. Rev. {\bf 129}, 647 (1963).

\bibitem{Ando} T. Ando, A. B. Fowler and F. Stern,
Rev. Mod. Phys. {\bf 54}, 437 (1982).


\bibitem{Zudov2} M. A. Zudov, R. R. Du, J. A. Simmons, and J. R. Reno,
Phys. Rev. B {\bf 64}, 201311(R) (2001).


\bibitem{Xie} Juren Shi and X. C. Xie, Phys. Rev. Lett. {\bf 91},
086801 (2003).




\bibitem{comment_resonance}
It is likely that
this sharp resonance will be rounded 
by a damping effect of impurity scattering.
But, still, we expect a large enhancement 
of other harmonic oscillations near $\omega \simeq \omega_c$.

\bibitem{Zudov3} M. A. Zudov, Phys. Rev. B {\bf 69}, 041304(R) (2004).


\bibitem{Sachdev} Adam C. Durst, Subir Sachdev, N. Read, and S. M. Girvin,
Phys. Rev. Lett. {\bf 91}, 086803 (2003).




\bibitem{R_xy} R. G. Mani {\it et al.}, Phys. Rev. B {\bf 69}, 161306 (2004).



\bibitem{CS1} A. Lopez and E. Fradkin, Phys. Rev. B {\bf 44}, 5246
(1991).


\bibitem{CS2} V. Kalmeyer and S.-C. Zhang, Phys. Rev. B {\bf 46}, 9889
(1992).

\bibitem{HLR} B. I. Halperin, P. A. Lee, and N. Read, 
Phys. Rev. B {\bf 47}, 7312 (1993).


\bibitem{comment_rhoxx} Composite fermions at $B^*=0$ are more affected by 
impurity scattering 
than electrons at $B=0$ because fluctuations in electron density (as a result 
of screening the impurity potential) generate fluctuating Chern-Simons 
gauge-field vector potential,
which may explain why $\rho_{xx}$ at $\nu=1/2$ is much larger 
than at $B=0$ for the same sample \cite{CS2,HLR}.
However, this scattering effect will just renormalizes the density of
states so that the amplitude of oscillation 
($\lambda$ in Eq.(\ref{dos}) in the text) 
in the density of states may be reduced somewhat, while leaving the main 
physical effect of radiation-induced magnetoresistance oscillation intact. 
As a corollary,
a higher mobility is necessary for radiation-induced 
magnetoresistance oscillation near $\nu=1/2$, if other parameters such as 
temperature are the same as those of electron system.


\bibitem{Kohn} For example, see L. Jacak, P. Hawrylak, and A. W\'{o}js, 
{\it Quantum Dots} (Springer, Berlin, 1998).


\bibitem{Du} R. R. Du {\it et al.}, Phys. Rev. Lett. {\bf 70}, 2944 (1993).

\bibitem{JainCFmass} J. K. Jain and R. K. Kamilla,
Phys. Rev. B {\bf 55}, R4895 (1997). 
 
\bibitem{comment_CFmass} It should be emphasized that
the CF mass does not have any dependence on 
the band mass of electron because 
the physics of lowest Landau level
has nothing to do with the band mass. 
In the text, the CF mass is compared with the electron band mass 
because the electron band mass is a convenient reference point.

\end{thebibliography}
\end{document}